\documentclass[letters, usenatbib]{mnras}
\usepackage{amssymb,amsmath}
\usepackage{graphicx}
\usepackage{xcolor,natbib}
\usepackage{multirow}
\usepackage[T1]{fontenc}
\usepackage{ae,aecompl}
\usepackage{newtxtext, newtxmath}
\usepackage{float}
\usepackage{subfigure}
\usepackage{longtable}
\usepackage{enumitem}
\usepackage{longtable,listings}
\usepackage[flushleft]{threeparttable}
\usepackage{parcolumns}
\usepackage{makecell}

\def\oc3{[O~{\sc iii}]$_c$}
\def\ob3{[O~{\sc iii}]$_b$}
\def\obj{SDSS J1617+0638}

\title[A normal BLAGN harboring a TDE]{A normal broad line AGN SDSS J1617+0638 as the host galaxy of a central tidal disruption event}

\author[Zhang]
{Xue-Guang Zhang \\
Guangxi Key Laboratory for Relativistic Astrophysics, School of Physical Science and Technology,
GuangXi University, Nanning, 530004, P. R. China}

\date{}

\begin{document}

\pagerange{\pageref{firstpage}--\pageref{lastpage}} \pubyear{2024}
\maketitle
\label{firstpage}

\begin{abstract} 
In this manuscript, strong clues are reported to support the normal broad line AGN \obj as the host galaxy harboring a central 
tidal disruption event (TDE). Through the optical flare in the CSS 8.5years-long light curve and the none-variability in the 
up-to-date ASAS-SN light curves, the theoretical TDE model described by the MOSFIT code can be applied in \obj. Meanwhile, 
considering the assumed central TDE expected continuum emissions not strong enough to describe the continuum emissions in the 
SDSS spectrum of \obj, an additional power law component from pre-existing AGN activity should be necessary in \obj. Furthermore, 
considering the short time duration to the observed date for the SDSS spectrum from the starting time of the assumed central TDE 
in \obj, TDE model expected accreting mass only about 0.03$M_\odot$ can lead to few effects of TDEs debris on the observed broad 
emission lines in the SDSS spectrum of \obj, indicating the TDE model determined BH mass simply consistent with the virial BH 
mass by broad emission lines, as determined results in \obj. Therefore, through both the photometric variability and the 
spectroscopic results, a central TDE can be preferred in the normal broad line AGN \obj~ with pre-existing central AGN activity 
and pre-existing broad emission line regions.  
\end{abstract}


\section{Introduction}

	TDEs (Tidal Disruption Events) as good beacons to massive black holes (BHs) and BH accreting systems have been well studied 
for more than five decades \citep{re88, np92, lu97, gm06, lr11, ce12, gr13, gm14, wz17, wy18, mg19, tc19, ry20, lo21, vg21, sg21, 
zl21, zh22, rk23, yr23, zh23a, ss24, tm24}, leading to apparent time-dependent variability patterns by accreting fallback debris 
from stars tidally disrupted by central BHs, when stars travelling through central BHs with distance smaller than tidal disruption 
radii but larger than event horizons. Based on TDEs expected variability patterns in different wavelength bands, so-far more than 150 
TDE candidates have been detected. More recent reviews on TDEs can be found in \citet{st19, gs21}.

	Among the reported TDE candidates, especially low-redshift optical TDE candidates, almost all optical TDE candidates are 
detected in quiescent galaxies. There are only a few TDE candidates reported in active galactic nuclei (AGN), such as in \citet{cm15, 
md15, bn17, sw18, yx18, am20, ll20, zs22, zh22b, sr23, zh24} for TDEs candidates in active galaxies or in changing-look AGN. 
Meanwhile, besides the host galaxy properties of TDE candidates, apparent broad Balmer and/or Helium emission lines related to 
TDEs debris are fundamental spectroscopic characteristics, such as the spectroscopic results in \citet{ve11, gs12, ht14, ht16, 
ht16b, bl17, lz17, gr19, ht19, sn20, hf20, hi21, ht23, zh24}. Spectroscopic properties of both broad emission lines and continuum 
emissions related to central BH accreting process can be applied as apparent signs for broad line AGN, however, those spectroscopic 
properties can also be expected from assumed central TDEs in host galaxies. Therefore, to report a central TDE in a definitely 
normal broad line AGN with pre-existing AGN activity is still a challenge.

\begin{figure*}
\centering\includegraphics[width = 18cm,height=8cm]{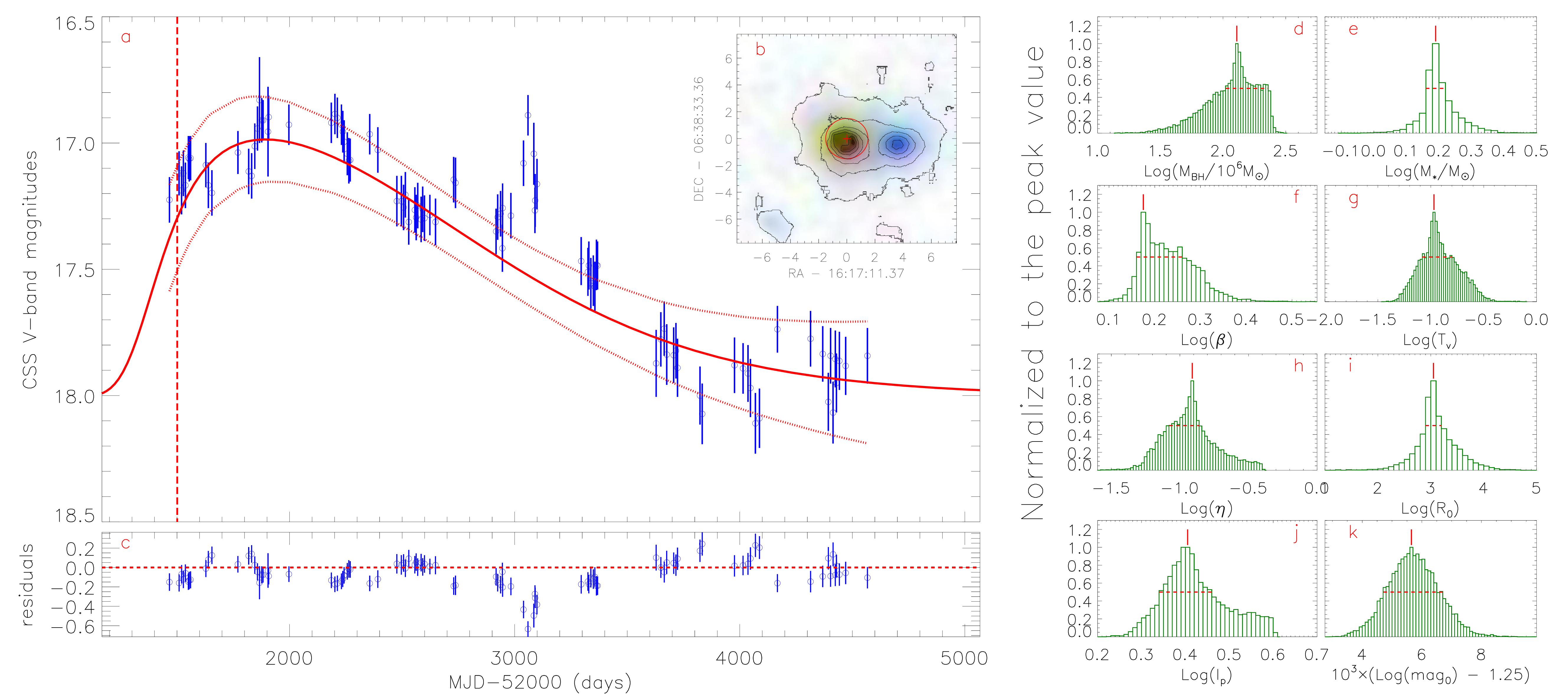}
\caption{Panel a shows the CSS light curve (open circles plus error bars in blue) in observer frame and the theoretical TDE model 
determined the best descriptions (solid red line) and the F-test technique determined corresponding 5$\sigma$ confidence bands 
(dotted red lines). In panel a, the vertical red dashed line marks the position MJD=53501 (the date for the SDSS spectrum of \obj). 
Panel b shows the photometric image (cut from SDSS) and the corresponding brightness contour for the \obj~ marked by a red circle 
(diameter 3\arcsec) with the central red cross as the central position, also for a near companion galaxy about 4\arcsec~ separated 
from the \obj. Panel c shows the residuals calculated by the light curve minus the best descriptions, with the horizontal red dashed 
line as residuals equal 0. The right panels show the MCMC determined posterior distributions of the TDE model parameters. In each 
right panel, the vertical red line marks the accepted value of the corresponding parameter, the horizontal red dashed line marks 
the upper and lower boundaries of the corresponding $1\sigma$ uncertainties of the model parameter.}
\label{lmc}
\end{figure*}

	In this manuscript, combining the unique variability properties of long-term light curves from the CSS (Catalina Sky Survey) 
\citep{dr09, dg14, gd17, sb21} and the ASAS-SN (All-Sky Automated Survey for Supernovae) \citep{sp14, ks17} with the spectroscopic 
properties of SDSS (Sloan Digital Sky Survey) spectrum observed around the times for an assumed TDE in SDSS pipeline classified 
broad line AGN, it will be a good try to find a TDE candidate harboured in a normal broad line AGN, based on the following two points. 
First, TDEs expected variability patterns can be detected in long-term light curves in a broad line AGN in SDSS. Second, besides 
TDE model expected continuum emissions, unique continuum emissions related to normal broad line AGN can be detected in the spectrum 
of the host galaxy.



	This manuscript is organized as follows. Section 2 presents the long-term photometric variability of \obj~ (=SDSS 
J161711.4+063833.5) at $z\sim0.229$ through the CSS and ASAS-SN, and the theoretical TDE model determined best descriptions. Section 
3 gives the discussions on spectroscopic properties probably related to normal broad line AGN. Section 4 gives final conclusions. 
And in this manuscript, we have adopted the cosmological parameters of $H_{0}=70{\rm km\cdot s}^{-1}{\rm Mpc}^{-1}$, 
$\Omega_{\Lambda}=0.7$ and $\Omega_{\rm m}=0.3$.

\begin{table*}
\caption{Model parameters for the central assumed TDE}
\begin{tabular}{ccccccccc}
\hline\hline
$\log(\frac{M_{\rm BH}}{\rm 10^6M_\odot})$ & $\log(\frac{M_\star}{\rm M_\odot})$  &  $\frac{R_\star}{\rm R_\odot}$
	& $\log(\beta)$ &
	$\log(T_{vis})$ & $\log(\epsilon)$ &  $\log(R_0)$ & $\log(l_p)$ & $mag_0$ \\
	\hline
	$2.11_{-0.09}^{+0.25}$  & $0.19_{-0.03}^{+0.03}$ & $1.87_{-0.11}^{+0.07}$ & $0.18_{-0.02}^{+0.08}$ &
	$-0.97_{-0.11}^{+0.16}$ & $-0.91_{-0.16}^{+0.05}$ & $3.05_{-0.15}^{+0.15}$ & $0.41_{-0.07}^{+0.05}$ &
	$17.98_{-0.02}^{+0.03}$ \\
	\hline\hline
\end{tabular}
\end{table*}

\section{Long-term Light curves of \obj~ from CSS and ASAS-SN}

	\obj~ is collected as the subject, due to the following three points. First, based on the long-term CSS light curve, 
there are TDE model expected variability properties. Second, based on the none variability properties of the long-term light 
curves from the ASAS-SN, variability properties of the TDE expected flare in CSS light curve are unique enough. Third, based 
on the SDSS spectrum observed around the time for the peak brightness of the optical photometric variability, the expected 
spectral energy distributions from the assumed central TDE can be determined, leading to further discussions on spectroscopic 
continuum emissions in \obj.

	The 8.5years-long light curve of \obj~ in panel a of Fig.~\ref{lmc} can be collected from the CSS with MJD from 53466 (Apr. 
2005) to 56567 (Oct. 2013), with a smooth decline trend which will be described by the following discussed theoretical TDE model. 
Moreover, as the shown photometric image of the \obj~ (cut from SDSS) in the panel b, there is a companion galaxy about 4\arcsec~ 
separated from the \obj. However, considering the SDSS gri-band magnitudes, the companion galaxy with magnitudes about [20.5,~19.0,
~18.5] is at least 2 magnitudes fainter than the \obj~ with magnitudes about [16.9,~16.9,~16.7], quite fainter than the shown CSS 
light curve. Meanwhile, after searching in the CSS, there are no light curves for the companion galaxy with searching radius 
about 4\arcsec. Therefore, the CSS light curve is for the \obj, not for the companion galaxy.

	It is interesting to check whether a TDE can lead to the optical flare in \obj. More detailed descriptions on the standard 
theoretical TDE model can be found in \citet{gr13, gm14, mg19}. Based on the public codes of TDEFIT/MOSFIT \citep{gm14, mg19}, the 
TDE model can be applied as follows based on templates of viscous-delayed accretion rates, very similar as what we have recently 
done in \citet{zh22, zh22b, zh22c, zh23a, zh23b, zh24}.

	Based on the TDEFIT/MOSFIT code provided time dependent fallback material rate $\dot{M}_{f}$ for standard TDEs cases with 
a solar-like main-sequence star (stellar mass $M_*=1M_\odot$) disrupted by a supermassive black hole (SMBH) ($M_{BH}=10^6M_\odot$) 
with different impact parameters $\beta$, templates of the viscous-delayed accretion rate $\dot{M}_{at}$ can be created by 
different viscous-delay time scales ($T_{vis}$)
\begin{equation}
\dot{M}_{a}(T_{vis})~=~\frac{exp(-t/T_{vis})}{T_{vis}}\int_{0}^{t}exp(t'/T_{vis})\dot{M}_{f}dt'
\end{equation}.
For TDEs with $M_{\rm BH}$ and $M_{*}$ different from $10^6{\rm M_\odot}$ and $1{\rm M_\odot}$, as discussed in \citet{gr13, mg19}, 
actual viscous-delayed accretion rates $\dot{M}$ and the corresponding time information are created by 
\begin{equation}
\begin{split}
	&\dot{M} = (\frac{M_{\rm BH}}{\rm 10^6M_\odot})^{-0.5}\times (\frac{M_{\star}}{\rm M_\odot})^2\times
	(\frac{R_{\star}}{\rm R_\odot})^{-1.5}\times\dot{M}_{a}(T_{vis}, \beta) \\
	&t = (1+z)\times (\frac{M_{\rm BH}}{\rm 10^6M_\odot})^{0.5}\times (\frac{M_{\star}}{\rm M_\odot})^{-1}\times
	(\frac{R_{\star}}{\rm R_\odot})^{1.5} \times t_{a}(T_{vis}, \beta)
\end{split}
\end{equation},
with $M_{\rm BH}$, $M_{\star}$, $R_{\star}$ and $z$ as central BH mass in units of ${\rm M_\odot}$, stellar mass in units of 
${\rm M_\odot}$, stellar radius in units of ${\rm R_{\odot}}$ and redshift of the host galaxy, respectively. And 
the more recent mass-radius relation discussed in \citet{ek18} has been accepted to determine $R_{\star}$ by giving $M_{\star}$ 
for main-sequence stars, to reduce one free model parameter.

	Then, based on the expected time-dependent actual accretion rates $\dot{M}$, the time-dependent output emission spectrum 
in observer frame can be determined by the simple blackbody photosphere model as discussed in 
\citet{gm14, mg19}, 
\begin{equation}
\begin{split}
&F_\lambda(t)=\frac{2\pi Gc^2}{\lambda^5}\frac{1}{exp(hc/(k\lambda T_p(t)))-1}(\frac{R_p(t)}{D})^2 \\
&R_p(t) = R_0\times a_p(\frac{\epsilon\dot{M}c^2}{1.3\times10^{38}M_{\rm BH}/{\rm M_\odot}})^{l_p} \\ 
&T_p(t)=(\frac{\epsilon\dot{M}c^2}{4\pi\sigma_{SB}R_p^2})^{1/4} \ \ \ \ \ \ a_p = (G M_{\rm BH}\times (\frac{t_p}{\pi})^2)^{1/3}	
\end{split}
\end{equation}
with $D$ as the distance to the earth calculated by the redshift $z$, $k$ as the Boltzmann constant, $T_p(t)$ and $R_p(t)$ 
represent the time-dependent effective temperature and radius of the photosphere, respectively, and $\epsilon$ as the energy 
transfer efficiency smaller than 0.4, $\sigma_{SB}$ as the Stefan-Boltzmann constant, and $t_p$ as time of the peak accretion.

	Based on the model expected emission spectrum $F_\lambda(t)$ convoluted with the transmission curves of the Johnson 
V-band filter, the time-dependent variability of the apparent CSS V-band magnitudes $mag(t)$ of \obj~ can be described by the model 
with eight parameters (redshift $z=0.229$ accepted to \obj) of the central BH mass $M_{\rm BH}$, the stellar mass $M_{\star}$ and 
the polytropic index $\gamma$ (4/3 or 5/3) of the disrupted main-sequence star, the impact parameter $\beta$, the viscous-delay 
time $T_{vis}$, the energy transfer efficiency $\epsilon$, the two parameters of $R_0$ and $l_p$ (to show the power law dependence 
of photosphere radius on the luminosity in the equation (3)) related to the blackbody photosphere, and an additional 
parameter $mag_0$ applied to describe non-variability component not related to the assumed TDE. Furthermore, through applications 
of the theoretical TDE model described by the MOSFIT code, there is only one restriction that the determined tidal radius by the 
model parameters is not smaller than event horizon of central BH.

\begin{figure*}
\centering\includegraphics[width = 18cm,height=3.75cm]{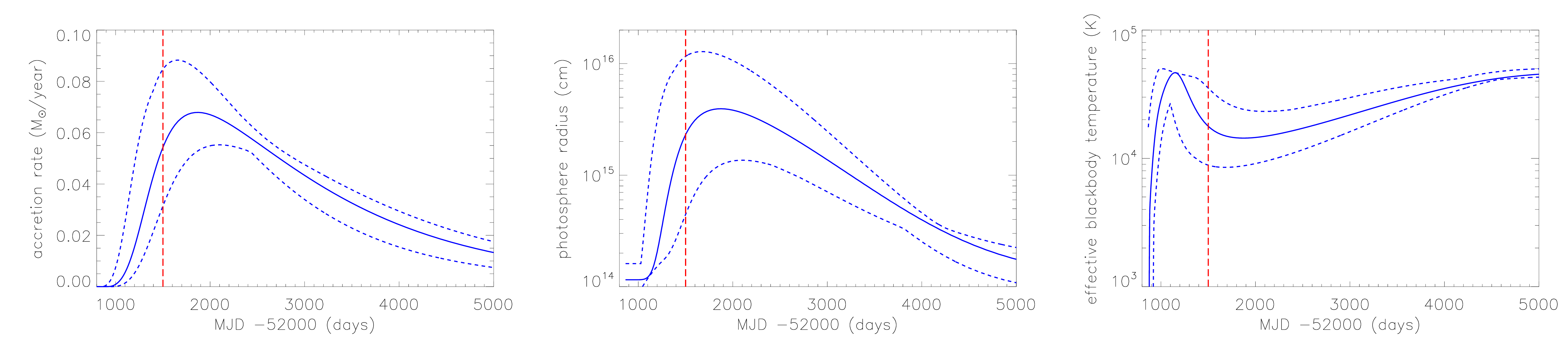}
\caption{Properties of the TDE model expected time dependent accretion rate $\dot{M}/(M_\odot/year)$ (left panel), photosphere 
radius $R_{p}$ (middle panel) and effective blackbody temperature $T_p$ (right panel) of \obj. In each panel, dashed lines in 
blue show the corresponding confidence band after accepted the $1\sigma$ uncertainties of the model parameters of \obj. In each 
panel, the vertical red dashed line marks the position of MJD=53501 (the MJD for the SDSS spectrum of \obj.}
\label{bdis}
\end{figure*}

\begin{figure}
\centering\includegraphics[width = 8cm,height=5cm]{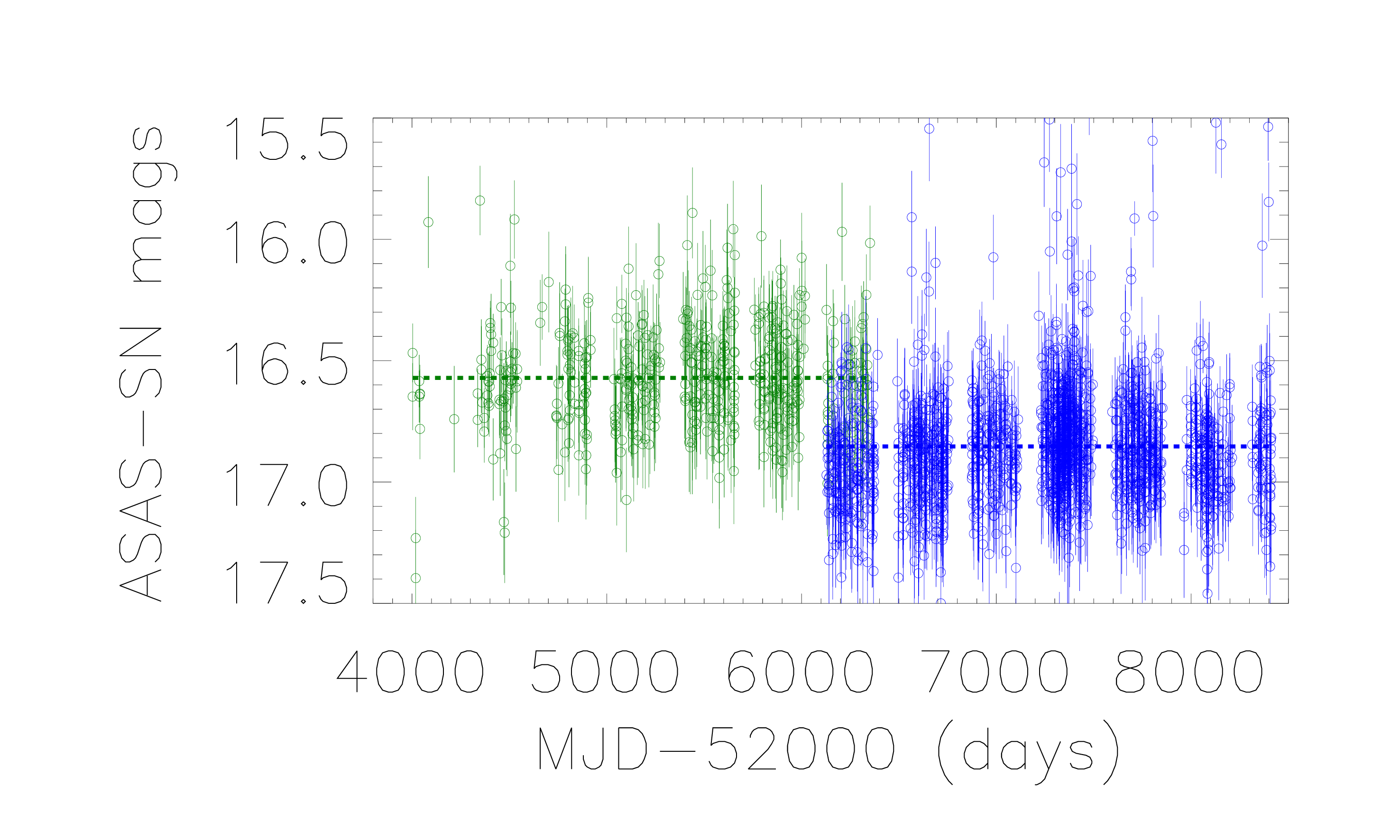}
\caption{The ASAS-SN V/g-band light curves (symbols in dark green and in blue) of \obj. Horizontal dashed lines in dark green 
and in blue mark the mean magnitudes of the V/g-band light curves.}
\label{as}
\end{figure}

\begin{figure*}
\centering\includegraphics[width = 18cm,height=3.5cm]{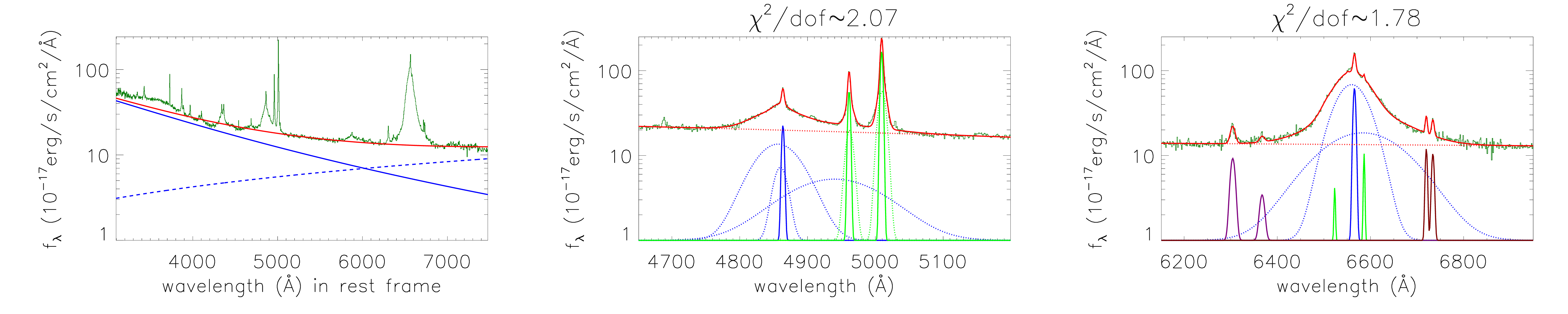}
\caption{Left panel shows the SDSS spectrum of \obj~ with PLATE-MJD-FIBERID=1732-53501-0251. In left panel, solid blue line shows 
the TDE model determined continuum emissions from the assumed blackbody photosphere with effective blackbody temperature 
$1.76\times10^4$K and photosphere radius $2.29\times10^{15}$cm, dashed blue line shows the determined power law component, solid 
red line shows the combinations of the TDE model determined continuum emissions and the power law continuum emissions. Middle panel 
and right panel show the best fitting results (solid red line) to the emission lines around the H$\beta$ and the H$\alpha$. In 
middle and right panel, dotted red line shows the determined continuum emission component, dotted blue lines and solid blue line 
show the determined Gaussian components in the broad and narrow Balmer line. In middle panel, solid and dotted green lines show 
the determined core and extended components in the [O~{\sc iii}] doublet. In right panel, solid lines in purple, in dark red show 
the determined components in [O~{\sc i}]$\lambda6300,6363$\AA~ and [S~{\sc ii}]$\lambda6716,6731$\AA~ doublets. In order to show 
clearer features of the determined continuum emission components and emission components in lines, the y-axis is plotted in 
logarithmic coordinate.}
\label{spec}
\end{figure*}

	Through the Maximum Likelihood method combining with the MCMC (Markov Chain Monte Carlo) technique \citep{fh13}, the 
best descriptions and the corresponding $5\sigma$ confidence bands can be determined to the CSS light curve by the discussed TDE 
model with the accepted uniform prior distributions of the model parameters, as shown in panel a of Fig.~\ref{lmc}. Meanwhile, 
the MCMC determined posterior distributions of the model parameters are shown in the right panels of Fig.~\ref{lmc}, leading the 
determined model parameters ($\gamma=4/3$) and the corresponding $1\sigma$ uncertainties to be listed in Table~1. Here, the 
parameter of stellar radius is determined by the mass-radius relation. And, the uncertainties of each model parameter are simply 
determined by the full width at half maximum of the corresponding posterior distribution of the parameter as shown in the right 
panels of Fig.~\ref{lmc}. The determined tidal disruption radius is about 1.49 times of the event horizon of the central BH. 
The model determined BH mass is about $128\times10^6{\rm M_\odot}$ in \obj, certainly smaller than the corresponding Hills limit 
value ${\rm 230\times10^6M_\odot}$ through the Equation (6) in \citet{yr23} by the determined stellar mass and stellar 
radius of the disrupted star in \obj, indicating the determined BH mass can be reasonably accepted in \obj.

	Before proceeding further, two points should be noted. First, the circularizations in TDEs as discussed in \citet{kc94, 
br16, hs16, zo20, lo21} are not considered in this manuscript. The circularization emissions in TDEs have been probably detected 
in TDE candidate AT2019avd in \citet{cd22}, due to the two clear peaks (or two clear phases) detected in the NUV and/or optical 
band light curves. However, among the more than 150 reported TDE candidates, there are rare TDE candidates of which optical 
light curves have re-brightened peaks, indicating the ratio of TDEs with clear circularization emissions is very low. Therefore, 
we mainly consider the simple case that the fallback timescales of the circularizations are significantly smaller than the 
viscous timescales of the accretion processes, and the fallback materials will circularize into a disk as soon as possible. 
Second, as shown in panel a of Fig.~\ref{lmc}, there is a secondary flare around MJD-52000=3150 with a very shorter 
time duration (about 100-200days) than that of the full TDE expected variability. However, considering the probable circularizations 
in the TDE candidates of AT2019avd in \citet{cd22} and ASASSN-15lh in \citet{lf16}, similar timescales of circularizations as those 
of the delayed accretion process could be expected. Therefore, not the circularizations but probable accretion instability 
\citep{cn15} are preferred to explain the secondary flare.

	Moreover, in order to test spectroscopic results of \obj~ in the next section through the simple blackbody photosphere 
model, the time dependent viscous-delayed accretion rate $\dot{M}(t)$ (corresponding to time dependent bolometric luminosity), 
the time dependent effective blackbody temperature $T_p(t)$ and photosphere radius $R_{p}(t)$ in \obj~ estimated by equation (3) 
are shown in Fig.~\ref{bdis}. Here, we should note that the applied model parameters are determined through the single-band light 
curve of \obj, probably leading to the determined model parameters being weakly constrained. Unfortunately, we can not 
find multi-band light curves around the optical flare in the CSS for the \obj, therefore, we directly applied the MCMC technique 
determined model parameters to determine the results in Fig.~\ref{bdis}. Comparing with the results in \citet{mg19} (see their Fig.~8), 
the time dependent $R_{p}$ and $T_p$ in \obj~ are common, especially the $R_{p}(t)$ mainly around $10^{15}$cm and $T_p(t)$ mainly 
around $10^4$K, indicating our determined results in \obj~ through TDE model are common and reasonable. Furthermore, assumed the 
central TDE in \obj, based on the $\dot{M}(t)$ in left panel of Fig.~\ref{bdis}, debris of about 0.48$M_\odot$ were accreted into 
the central BH of \obj. Moreover, as what will be discussed in the next section, the SDSS spectrum of \obj~ was observed at 
MJD=53501, about 500days from the starting time of the optical flare, indicating only about 0.03$M_\odot$ have been accreted into 
the central BH of \obj~ before MJD=53501. The small accreting mass before MJD=53501 should indicate few effects of TDE debris on 
probably pre-existing BLRs (broad emission line regions) in \obj~ as a potential normal broad line AGN, especially after considering 
several to hundreds of solar masses as total mass of BLRs in broad line AGN as discussed in \citet{pe97, bf03}.

	Furthermore, the follow-up up-to-date long-term V/g-band light curves of \obj~ can be collected from the ASAS-SN with 
MJD from 56003 (Mar. 2011) to 60411 (Apr. 2024), and shown in Fig.~\ref{as}. Here, we should note that the ASAS-SN light curves of 
\obj~ are much brighter than the CSS light curves. After carefully checking the photometric image cut from the ASAS-SN, 
it can be confirmed that the shown light curves in Fig.~\ref{as} are really for the \obj~ but probably including contributions of 
the companion galaxy due to blended images. Unfortunately, considering the V-band magnitude about 19.7 of the companion galaxy, 
even combining with the contributions of the \obj~ (V-band magnitude about 17.9, the mean value of data points in the CSS V-band 
light curve after MJD-52000 larger than 4000), the expected V-band magnitude should be 17.56\ in the ASAS-SN, still 1 magnitudes 
fainter than the ASAS-SN V-band magnitude of \obj. Although we do not know the real reason why the ASAS-SN provided light curves 
of \obj~ are so brighter, we can safely confirm none apparent variability in the ASAS-SN light curves of \obj. Therefore, the 
optical flare shown in Fig.~\ref{lmc} can be confirmed to be unique enough, and be very different from the intrinsic AGN variability 
as common features in different epochs.

\section{Spectroscopic results of \obj}

	The SDSS spectrum of \obj~ with PLATE-MJD-FIBERID=1732-53501-0251 is collected from SDSS DR16 \citep{ap20} and shown in 
left panel of Fig.~\ref{spec}, with $z=0.229$ determined through the abundant emission lines. Moreover, as shown in panel b of 
Fig~\ref{lmc}, the spectrum from the SDSS fiber (diameter 3\arcsec) for the \obj~ have few contributions from the companion galaxy. 
Then, properties of both the spectral energy distributions (SEDs) (or continuum emission properties) and the emission lines 
(especially broad emission lines) are analyzed as follows.

	Based on the determined effective blackbody temperature $1.76\times10^4$K and photosphere radius $2.29\times10^{15}$cm 
tightly related to the assumed central TDE in \obj~ at MJD=53501, the corresponding continuum emissions can be determined and 
shown as solid blue line in left panel of Fig.~\ref{spec}. It is clear that the TDE determined continuum emissions are not 
consistent with the V-band (4600\AA~ to 7400\AA~ in observer frame, 3742\AA~ to 6021\AA~ in rest frame) SEDs of the SDSS spectrum 
of \obj~ at MJD=53501, and there should be an additional continuum emission component to be considered. After accepted the TDE 
model determined continuum emission component, an additional power law function is considered and shown as dashed blue line 
($\propto~5.68\lambda^{1.21\pm0.03}$) in left panel of Fig.~\ref{spec}, in order to find the better descriptions to the SEDs 
(especially V-band) of \obj.

	Although combining a TDE expected continuum emission component with an additional power law component can be applied 
to describe the observed SEDs of \obj~ at MJD=53501, it is necessary to check the physical origin of the additional power law 
component, from host galaxy or from central pre-existing AGN activity. If the power law component was from host galaxy, due to 
the redness of the power law component, apparent absorption features from old stellar populations should be expected, however 
as shown in the left panel of Fig.~\ref{spec}, there are no apparent absorption features. Here, similar as what have recently 
done in \citet{zh24b}, we have actually applied the Simple Stellar Population method \citep{ref33, ref32} to determine probable 
contributions of stellar light. However, due to none absorption features, the contributions of stellar 
light can be negligible. Therefore, the power law component $\lambda^{\sim1.21}$ should be not from the quiescent host galaxy 
of \obj. In other words, it is preferred that the additional power law component should be not from stellar populations but related 
to pre-existing central AGN activity.


	Besides the properties of SEDs of \obj~ at MJD=53501, it is necessary to discuss properties of broad Balmer emission 
lines of \obj. Similar as what we have recently done in \citet{zh21, zh23c, zh24}, the emission lines around H$\beta$ (within 
rest wavelength range from 4600\AA~ to 5250\AA) and around H$\alpha$ (within rest wavelength range from 6150\AA~ to 6950\AA) 
can be measured by multiple Gaussian functions, two broad Gaussian functions (second moment larger than 400km/s) plus one narrow 
Gaussian function (second moment smaller than 400km/s) for the broad and narrow H$\beta$ (H$\alpha$), two Gaussian functions 
for the extended and core components in each [O~{\sc iii}]$\lambda4959,5007$\AA~ line, two Gaussian functions for the 
[O~{\sc i}]$\lambda6300,6363$\AA~ doublet ([N~{\sc ii}]$\lambda6548,6583$\AA~ doublet, [S~{\sc ii}]$\lambda6716,6731$\AA~ doublet). 
Then, a power law function is applied to describe the continuum emissions underneath the emission lines around H$\beta$ (H$\alpha$). 
When the model functions are applied, only two criteria are accepted. First, the emission intensity of each Gaussian component 
is not smaller than zero. Second, the flux ratios of the [O~{\sc iii}]$\lambda4959,5007$\AA~ doublet 
([N~{\sc ii}]$\lambda6548,6583$\AA~ doublet) are fixed to be 1:3. Then, through the Levenberg-Marquardt least-squares minimization 
technique (the known MPFIT package) \citep{mc09}, the best fitting results to the emission lines are shown in the middle and 
right panel of Fig.~\ref{spec}, with $\chi^2/dof\sim2$. The broad H$\alpha$ (broad H$\beta$) can be described by two Gaussian 
components with model parameters of [6583.6$\pm$1.3, 86.8$\pm$1.5, 3797$\pm$87] (the first value as central wavelength in units 
of \AA, the second value as the second moment in units of \AA~ and the third value as the flux in units of 
${\rm 10^{-17}erg/s/cm^2}$) and [6559.7$\pm$0.3, 34.6$\pm$0.4, 5836$\pm$104] ([4857.6$\pm$0.5, 25.9$\pm$0.7, 871$\pm$33] and 
[4903.3$\pm$3.3, 83.2$\pm$2.6, 1185$\pm$52]).

	As shown in Section 2, total accreting mass about 0.03$M_\odot$ before MJD=53501 for the assumed TDE in \obj~ should 
lead broad line emissions having few contributions related to the TDE, indicating there should be consistency between the 
TDE model determined BH mass and the virial BH mass in \obj. As discussed in \citet{pf04, gh05, rh11, sh11} by applications 
of Virialization assumption to central BLRs, the virial BH mass of \obj~ can be estimated as 
$1.92_{-0.06}^{+0.14}\times10^8{\rm M_\odot}$ by  
\begin{equation}
\frac{M_{\rm BH}}{\rm M_\odot}~=~2.2\times10^6\times(\frac{L_{\rm H\alpha}}
	{\rm 10^{42}erg/s})^{0.56}\times(\frac{FWHM_{\rm H\alpha}}
	{\rm 1000km/s})^{2.06} 
\end{equation},
with $L_{\rm H\alpha}\sim(1.50\pm0.03)\times10^{43}$erg/s and $FWHM_{\rm H\alpha}\sim4235\pm80$km/s as the line luminosity and 
the full width at half maximum of the measured broad H$\alpha$ of \obj, respectively. And the uncertainties of the virial BH 
mass are determined by the uncertainties of the $L_{\rm H\alpha}$ and $FWHM_{\rm H\alpha}$. Considering the uncertainties of the 
virial BH mass and the TDE model determined BH mass (about $1.29_{-0.25}^{+1.01}\times10^8{\rm M_\odot}$), it can be roughly 
accepted that the two kinds of BH mass are simply consistent.

	Moreover, if the observed broad Balmer emission lines were totally related to the TDE debris in \obj~ as a quiescent 
galaxy, the expected outer distance at MJD=53501 of the broad line emission materials to the central BH as discussed in 
\citet{gr13} can be estimated as $R_{out}\sim2\times(\frac{G\times M_{\rm BH}\times t^2}{\pi^2})^{1/3}\sim11.4{\rm light-days}$. 
However, considering the continuum luminosity about $1.27\times10^{44}$erg/s at 5100\AA~ from the spectrum in left panel of 
Fig.~\ref{spec}, through the empirical R-L relation in \citet{bd13}, the estimated BLRs size is about 40light-days, about 
three times larger than the expected outer distance of the TDE debris. Therefore, indirect clues can be reported to support 
the pre-existing BLRs, otherwise there should be very different virial BH mass from the TDE model determined BH mass.

	Before ending of the manuscript, an additional point should be noted. If the results shown in left panel of Fig.~\ref{spec} 
were intrinsically true, there should be a pre-existing accretion disk related to central pre-existing AGN activity. Effects of 
the pre-existing accretion disk on TDEs properties can be simply discussed by the following three points. First, as more recent 
discussed in \citet{rm24}, lower mass density in pre-existing accretion disk should have tiny effects on properties of TDEs debris 
and then on TDEs expected variability patterns. Second, as the reported TDEs in AGNs in the literature as listed in Introduction, 
TDEs expected variability patterns can be detected in the long-term light curves. Third, as discussed in \citet{kb17, cp19, cp20}, 
after considering the evolutions of the TDEs debris modified by collisions with the pre-existing disk, the expected variability 
should be not very different from the expected results by the standard TDE model described by the MOSFIT code, once there were 
very weak collisions between the TDEs debris and the pre-existing accretion disk. Therefore, the TDE model expected variability 
pattern in \obj~ can be reasonably accepted in a normal broad line AGN with a pre-existing accretion disk.

\section{Main Summary and Conclusions}

	The objective of this manuscript is to report a TDE candidate in a normal broad line AGN \obj. Through the optical 
flare of \obj~ in the 8.5years-long CSS light curve and the following up-to-date ASAS-SN light curves without variability, 
a central TDE can be preferred in the \obj. Meanwhile, in order to describe the optical continuum emission properties in the 
SDSS spectrum, besides the TDE model expected component, an additional power law component from pre-existing central AGN activity 
should be necessary in \obj. Furthermore, considering short time duration between the observed date of the SDSS spectrum and 
the starting time for the optical flare, pre-existing BLRs related to central AGN activity should be also preferred in \obj. 
Through the spectroscopic results and the photometric variability, the \obj~ should be a remarkable candidate for a normal 
broad line AGN harboring a central TDE.

\section*{Acknowledgements}
Zhang gratefully acknowledge the anonymous referee for giving us constructive comments and suggestions to 
greatly improve our paper. Zhang gratefully acknowledges the kind financial support from GuangXi University, and the grant 
support from NSFC-12173020 and NSFC-12373014. This manuscript has made use of the data from the SDSS projects 
(\url{http://www.sdss3.org/}), and the public code of TDEFIT (\url{https://github.com/guillochon/tdefit}) and MOSFIT 
(\url{https://github.com/guillochon/mosfit}), and the MPFIT package (\url{http://cow.physics.wisc.edu/~craigm/idl/idl.html}), 
and the emcee package (\url{https://pypi.org/project/emcee/}). 

\section*{Data Availability}
The data underlying this article will be shared on reasonable request to the corresponding author
(\href{mailto:xgzhang@gxu.edu.cn}{xgzhang@gxu.edu.cn}).

\label{lastpage}
\end{document}